# Opening and reversible control of a wide energy gap in uniform monolayer graphene

*Cheolho Jeon[1], Hacheol Shin[1], Inkyung Song[1], Minkook Kim[1], Ji-Hoon Park[1], Jungho Nam[1],*

*Dong-Hwa Oh[1], Sunhee Woo[2], Chan-Cuk Hwang[3], Chong-Yun Park[1] & Joung Real Ahn[1,4]*

[1]Department of Physics, Sungkyunkwan University, Suwon 440-746, Republic of Korea,

[2]College of Pharmacy, Chungnam National University, Daejeon 305-764, Republic of Korea,

[3]Beamline Research Division, PAL, POSTECH, Pohang, 790-784, Republic of Korea,

[4]SAINT and Integrated Nanostructure Physics (CINAP), Institute for Basic Science(IBS),

Sungkyunkwan University, Suwon 440-746, Republic of Korea

Correspondence and requests for materials should be addressed to C.Y. Park (cypark@skku.edu) and J.R. Ahn (jrahn@skku.edu).

**ABSTRACT**

For graphene to be used in semiconductor applications, a 'wide energy gap' of at least 0.5 eV at the Dirac energy must be opened without the introduction of atomic defects. However, such a wide energy gap has not been realized in graphene, except in the cases of narrow, chemically terminated graphene nanostructures with inevitable edge defects. Here, we demonstrated that a wide energy gap of 0.74 eV, which is larger than that of germanium, could be opened in uniform monolayer graphene without the introduction of atomic defects into graphene. The wide energy gap was opened through the adsorption of self-assembled twisted sodium nanostrips. Furthermore, the energy gap was reversibly controllable through the alternate adsorption of sodium and oxygen. The opening of such a wide energy gap with minimal degradation of mobility could improve the applicability of graphene in semiconductor devices, which would result in a major advancement in graphene technology.

Graphene is a densely packed two-dimensional carbon crystal. Its structure is robust due to the planar $sp^2$ hybridizations of the carbon atoms. The electrons of the carbon atoms produce a half-filled band with a linear dispersion near the first Brillouin zone corners, which results in a zero-gap semiconductor. The linear dispersion around the Fermi energy ($E_F$) reflects massless Dirac fermion behavior[1]. The ballistic transport of electrons, which was expected from the massless Dirac fermion behavior, has been experimentally observed in graphene: the average propagation length without scattering is a few micrometers at room temperature (RT)[2].

Recent work has focused on the development of graphene-based semiconductor devices for the practical application of the massless Dirac fermion behavior[3-7]. Because graphene is a zero-gap semiconductor, an energy gap must be opened around the Dirac energy ($E_D$) to make graphene applicable in semiconductor devices. Herein, an energy gap refers to a gap around $E_D$. At least 0.5 eV is required to operate a graphene-based device at RT, and wider energy gaps would increase the on–off ratios of the devices. The mobility of graphene with an energy gap is inversely proportional to the magnitude of the energy gap, which means energy gaps greater than 1 eV are undesirable. The appropriate energy gap is between 0.5 and 1 eV. The mobility can be further reduced if atomic defects are created during the opening of an energy gap, as observed in hydrogenated graphene[8-10] and graphene oxides[11-12]. Because the opening of energy gaps between 0.5 and 1 eV without producing atomic defects or chemically terminated structures in the graphene is difficult, graphene-based devices have not been produced with suitable on–off ratios and mobilities at RT.

An energy gap around $E_D$ can be opened when the sublattice symmetry of two equivalent carbon atoms in the unit cell of the monolayer graphene is broken[1]. This mechanism allows wide energy gaps of a few eV to open when graphene forms strong chemical bonds with molecules such as hydrogen or oxygen[13-15]. However, the strong chemical bonds act as atomic defects and diminish the Dirac fermion behavior of the graphene, which significantly reduces the mobility. Quantum confinement in graphene nanoribbons[1,16,17], nanomeshes[18,19], or nanodots[20,21] with chemically terminated edge structures can also open wide energy gaps. These graphene nanostructures must be less than 2 nm for energy gaps greater than 0.5 eV to be opened[22]. Such structures are too small to be produced by lithography without the introduction of edge defects that would degrade the mobility. This work reports angle-resolved photoemission spectroscopy (ARPES) and the scanning tunneling microscopy (STM) evidence of a wide energy gap of 0.76 eV. The gap was opened at RT in uniform monolayer graphene with single domain sizes of at least 150 nm, and no atomic defects or chemically terminated edge structures were created in the graphene. The wide energy gap was reversibly controllable at RT without further thermal treatments. Self-assembled twisted sodium nanostrips were used to form notably weak chemical bonds in the graphene, which could minimally affect the mobility after the wide energy gap was opened.

**RESULTS**

**ARPES experiments of the opening of a wide energy gap at the Dirac energy. Figure 1** shows the electronic band structure of the monolayer graphene with increasing sodium coverage. An energy gap at $E_D$ was measured from the energy distribution curves (EDCs) (**Figure 1b** and Supplementary **Figure S2a**). The n-type monolayer graphene exhibited an $E_D$ value of -0.35 eV, with an energy gap of 0.20-0.28 eV (**Figure 1b**). Notably, the existence (or absence) of the energy gap at the $E_D$ value of monolayer graphene is still a topic of debate[23,24]. S. Y. Zhou *et al.* reported an energy gap at $E_D$, which they ascribed to a symmetry breaking induced by a buffer layer located between the monolayer graphene and an SiC substrate[23]. In contrast, E. Rotenberg *et al.* suggested that an energy gap at the $E_D$ value of the monolayer graphene did not exist and described an electronic band structure at $E_D$ in terms of a many-body interaction[24]. For the initial sodium coverage, a new energy band appeared with an $E_D$ value of -1.2 eV (**Figures 1c-e**). Its photoemission intensity increased with increasing sodium coverage, whereas the photoemission intensity of the energy band of the pristine graphene became weaker. Interestingly, the $E_D$ values of both energy bands varied slightly with increasing sodium coverage. The momentum distribution curves (MDCs) at $E_F$ (**Figure 1h**) show that the Fermi momentums of both energy bands were almost maintained with respect to the sodium coverage; only their photoemission intensities changed. At higher sodium coverages, the energy band of the pristine graphene disappeared, leaving only the new energy band (**Figure 1h**), whose energy gap was 0.74 eV (**Figure 1f** and Supplementary **Figure S2b**). The Fermi velocities $V_F$ were extracted from the energy dispersions of the pristine graphene and the sodium-covered monolayer graphene. The $V_F$ value of the pristine monolayer graphene was $0.93\times10^6$ m/s, which is consistent with the previous reports[25,26]. When the wide energy gap of 0.74 eV was opened by the sodium adsorption, the $V_F$ slightly decreased to $0.79\times10^6$ m/s. This reduction suggested that the $V_F$ value of the pristine monolayer graphene was not substantially decreased by the sodium adsorption at RT. At saturation sodium coverage, the energy gap of the new band was slightly decreased to 0.65 eV (**Figure 1g** and Supplementary **Figure S2c**), which suggests that the optimum sodium coverage for the maximum energy gap of 0.74 eV was approximately half the saturation sodium coverage.

After the adsorption of sodium at RT, the sample was cooled to a low temperature (LT) of 60 K to clarify a thermal broadening effect on the wide energy gap opening. The ARPES intensity maps measured at 60 K (Supplementary **Figure S3**) did not change compared to those measured at RT, except for the spectral width. This result suggests that the wide energy gap opening was not affected by thermal broadening and that the wide energy gap at $E_D$ is intrinsic. According to an ARPES experiment of the sodium adsorption onto monolayer graphene at RT and 85 K reported by C. G. Hwang *et al.*[27], who measured the ARPES intensity maps along the Γ–K direction, half the energy bands of monolayer graphene were observed. Furthermore, because the constant-energy maps were not measured, the K point may be not determined accurately. Therefore, the existence of an energy gap at $E_D$ was not obvious and was not discussed. Bostwick *et al.* reported the ARPES intensity maps of a potassium-covered monolayer graphene, where potassium was adsorbed at 100 K[28], and noted that no energy gap opening at $E_D$ was observed. Thus, whether a wide energy gap was opened when sodium was adsorbed at a LT was not obvious. After sodium was adsorbed onto the pristine graphene at 60 K, the ARPES intensity maps were measured at 60 K (Supplementary **Figure S4**). In contrast to the RT adsorption of sodium, the electronic band structure of the pristine graphene shifted rigidly toward the higher-binding-energy side, and a new energy band was not observed. This result leads to the conclusion that an energy gap at $E_D$ was not opened when sodium was adsorbed at LT, which is consistent with the results of Bostwick *et al.* This result also suggests that the wide energy gap opening at $E_D$ of the sodium-covered monolayer graphene is different from the debated small energy gap opening at $E_D$ of the pristine graphene. For a systematic experiment, ARPES intensity maps were measured after other alkali metals (potassium and cesium) were adsorbed onto the pristine graphene at RT (Supplementary **Figure S5**). A new energy band with a different $E_D$ was observed; however, unlike the sodium case, the electronic band structure at $E_D$ was slightly changed from that of the pristine graphene, and a wide energy gap opening was not observed. Thus, for potassium and cesium, the answer to the question of whether an energy gap at $E_D$ was opened at RT was not obvious because the small energy gap of the pristine graphene is still a subject of debate. This suggests that the wide energy gap opening induced by sodium adsorption at RT may originate from the surface morphology of sodium atoms at RT. Furthermore, when sodium was adsorbed at RT, the intensity of the $(6\sqrt{3} \times 6\sqrt{3})R30°$ low-energy electron diffraction (LEED) pattern of the pristine monolayer graphene was slightly diminished, and no additional ordered LEED pattern, such as a $(2 \times 2)$ LEED pattern, which was observed in other alkali metal experiments, was produced (Supplementary **Figure S6**). Such $(2 \times 2)$ superstructures cannot open a wide energy gap at the $E_D$[28-30], which also suggests that the wide energy gap opening at $E_D$ is due to the surface morphology of sodium atoms at RT.

**STM experiments of the surface morphology of sodium nanoribbons.** To better understand the wide energy gap opening, we acquired STM images (**Figure 2**). The STM images of the pristine graphene with domain widths greater than 150 nm show bright 6×6 superstructures, which originate from an underlying buffer layer with unsaturated carbon atoms (**Figures 2a and 2b**). The hexagonal carbon image of the monolayer graphene is also shown in the enlarged image in **Figure 2b**. Sodium initially produced triangular nanoclusters with diameters of approximately 3 nm (**Figure 2c**), which suggests that it was mobile at RT. With increased sodium coverage, the nanoclusters transformed into nanostrips with widths similar to the sizes of the nanoclusters (**Figure 2d**). The nanostrips were elongated and twisted as the sodium coverage increased, whereas their widths were maintained, which resulted in the first sodium overlayer that comprised closely packed twisted nanostrips (**Figure 2e**). Interestingly, the twisted nanostrip structure was maintained in the second sodium layer with saturation sodium coverage (**Figure**

**2f**). The STM images show two different graphene regions: those with the twisted sodium nanostrips and those without the nanostrips. This observation let us conclude that each of the energy bands observed in the ARPES spectra (**Figures 2g and 2h**) was related to a different graphene region shown in the STM images. The wide-gapped energy band with the lower $E_D$ was attributed to the graphene with twisted sodium nanostrips, and the narrow-gapped energy band with the higher $E_D$ was due to the pristine graphene without sodium. The sodium overlayer was saturated at the second layer, which made the optimal sodium coverage of the wide energy gap of 0.74 eV close to the full coverage of the first sodium layer, where the alkali metal coverage of the first layer was 0.25 monolayers[31]. Thus, the decrease in the energy gap at saturation sodium coverage was clearly related to the growth of the second sodium layer.

**ARPES experiments of the reversible control of the wide energy gap at the Dirac energy.** The sodium-covered monolayer graphene was exposed to oxygen at RT to assess whether the electronic band structure of the pristine graphene could be recovered (**Figure 3**). Exposure to 0.5 L of oxygen (1 L = $1 \times 10^{-6}$ Torr) resulted in an energy band with $E_D$ = 0.5 eV, similar to that of the pristine graphene, whereas the wide-gapped energy band was largely maintained (**Figure 3b**). An increase in the oxygen exposure resulted in increased photoemission intensity of the narrow-gapped energy band; however, the photoemission intensity of the wide-gapped energy band decreased (**Figures 3c and 3e**). At saturation, the electronic band structure of the pristine monolayer graphene was recovered (**Figure 3f**). Oxygen exposure reversed the energy gap opening induced by sodium. Because oxygen was not reactive with graphene at RT, it reacted with the sodium overlayer and produced a layer of sodium oxide. The valence electron of sodium transferred to oxygen instead of to graphene because of their different electronegativities. This transfer induced the oxygen to recover the electronic band structure of the pristine graphene. The reversible control of the energy gaps of graphene was demonstrated by the alternate adsorption of sodium and oxygen at RT (**Figure 4**). Sodium was adsorbed onto the sodium oxide layer, which resulted in an energy gap and an $E_D$ value similar to those of the sodium-adsorbed monolayer graphene (**Figure 4d**). Subsequent oxygen adsorption reproduced the electronic band structure of the first sodium oxide layer on the graphene (**Figure 4e**). The reversibility between the wide energy gapped graphene and the pristine graphene was maintained for up to three cycles of alternate sodium and oxygen adsorption. Further adsorption was possible; however, the photoelectron intensity of the graphene was too greatly diminished by the multiple sodium oxide layers. The sodium oxide layer prevents the transfer of the valence electron from the additional sodium to the underlying graphene. Therefore, additional sodium may diffuse through the boundaries of the first sodium oxide layer and situate between the first sodium oxide layer and graphene. In this case, additional oxygen may diffuse through the boundaries of the first sodium oxide layer to form an additional sodium oxide layer between the first sodium oxide layer and the graphene. The reversible control can be understood by the repetitive formation of sodium oxide layers.

**DISCUSSION**

Because the energy bands of the sodium overlayer do not exist at the $E_D$ at the K point, the possibility of hybridization between the energy bands of sodium and graphene can be excluded. A simple charge transfer from sodium cannot open an energy gap without breaking the sublattice symmetry, as shown by theoretical calculations of epitaxial graphene under an electric field[32]. A superstructure, which was not observed in the STM images or indicated by the low-energy electron diffraction results, can also be excluded. The sodium overlayer is different from other overlayer structures in its short-range incommensurability with graphene, its inhomogeneous charge distribution, and its randomly twisted quasi-1D structure. The incommensurability between gold and graphene on a Ru(0001) surface has been observed to open an energy gap of approximately 0.2 eV despite the weak interaction between gold and graphene[33]. However, the wide energy gap of 0.74 eV cannot be explained by only the incommensurability. Charge impurities were shown to induce backscattering in STM experiments on graphene on a Si oxide film[34], and the bound states were theoretically predicted to arise from single charge impurities[35]. The sodium overlayer resulted in randomly twisted quasi-1D charge impurities of finite width. If these quasi-1D charge impurities were periodically distributed, they could not produce a bound state, as shown by the Dirac–Kronig–Penny model[36]. Therefore, the random twisting of the quasi-1D charge impurities, which resulted in a disordered potential, could produce a bound state related to the wide energy gap opening. A combined mechanism of incommensurability and randomly twisted quasi-1D charge distribution may cause the wide energy gap opening. Theoretical reproduction of the wide energy gap by first principles or numerical calculations would be difficult due to the incommensurability and the non-uniformity of the sodium overlayer.

In summary, via the formation of self-assembled twisted sodium nanostrips on uniform epitaxial graphene, a wide energy gap of 0.74 eV at $E_D$ was opened in uniform graphene without chemically terminated edge structures. The energy gap was directly measured using ARPES, and the sodium nanostrips were observed using STM. The twisted sodium nanostrips, which caused non-uniform spatial charge distributions on the graphene, and their incommensurability with the graphene may have opened the wide energy gap. The interactions between sodium and graphene are much weaker than those in chemically terminated structures that involve hydrogen or oxygen; therefore, they may not reduce the mobility. The wide energy gap reverted to the original electronic band structure of the pristine graphene when the sodium was subsequently oxidized. The transition between the wide-energy-gapped graphene and the pristine graphene was reversibly controllable through the repeated adsorption of sodium and oxygen. This approach to produce wide-gapped graphene without chemically terminated edge structures may contribute to the development of graphene-based electronic devices that require both a high on–off ratio and high mobility.

**METHODS**

Epitaxial graphene was grown on an n-type 6H-SiC(0001) wafer (CREE) by annealing at 1470 K. The ARPES spectra were measured with a commercial angle-resolved photoelectron spectrometer (R3000, VG-Scienta) using monochromated He-II radiation ($hv$ = 40.8 eV, VG-Scienta) at RT. The base pressure was less than $5.0 \times 10^{-11}$ Torr, and the overall energy and angular resolutions were 70 meV and 0.1°, respectively. The ARPES spectra at LT were measured at the 4A2 beamline in the Pohang Accelerator Laboratory in Korea. For the ARPES measurements, the correct position of the K point in the first

Brillouin zone was determined from a Fermi surface map, which is a constant-energy map at $E_F$ (see Supplementary Information, Figure S1). The STM images were acquired using LT-STM (Omicron) at RT in vacuum with a pressure less than $2.0 \times 10^{-11}$ Torr. Sodium was deposited using a commercial getter source (SAES) at RT. The oxygen gas was introduced into the chambers using a variable leak valve at RT.

**ACKNOWLEDGMENTS**

This study was supported by the Priority Research Centers Program through the National Research Foundation of Korea (NRF) (2011-0031392) and the National Research Foundation of Korea (NRF) grant, which were funded by the Korean government (MEST) (No. 2012R1A1A2041241) and the Research Center Program of IBS (Institute for Basic Science) in Korea. The experiments at the PLS were supported in part by MEST and POSTECH.

**AUTHOR CONTRIBUTIONS**

CJ, HCS, and CCH performed photoemission spectroscopy experiments. IS, MKK, and JHN performed scanning tunneling microscopy experiments. DHO, CYP, and JRA contributed to the discussion of the results. CJ, SHW, and JRA wrote the manuscript. CJ and MKK prepared the figures.

**ADDITIONAL INFORMATION**

**Competing financial interests**: The authors declare no competing financial interests.

**FIGURES**

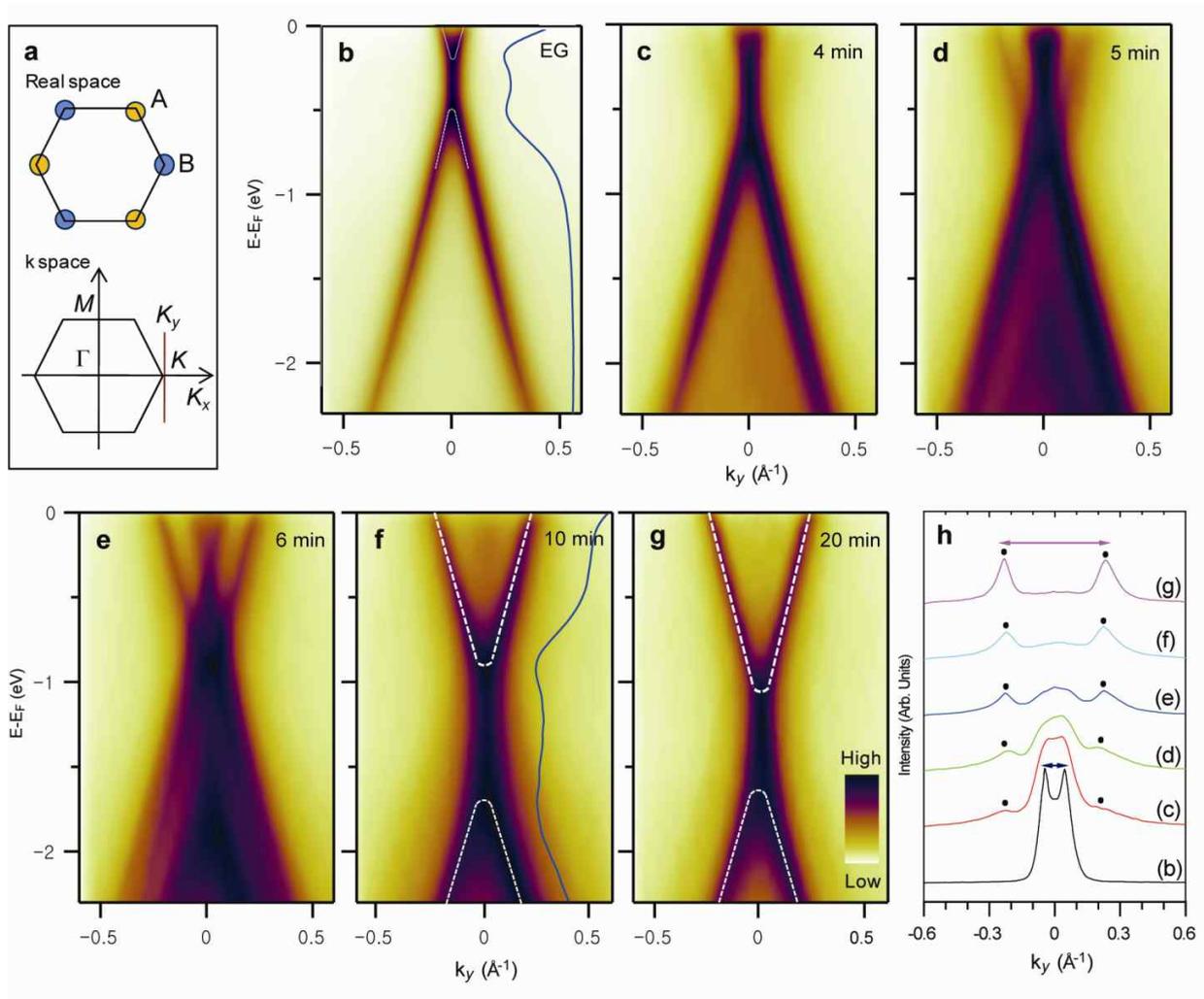

**Figure 1 | Variations of the electronic band structures of monolayer graphene near the K point with increasing sodium coverage**. (a) Atomic structure and first Brillouine zone of monolayer graphene. (b) ARPES intensity map of the pristine monolayer graphene along the $k_y$ direction. (c-g) ARPES intensity maps of the sodium-adsorbed monolayer graphene along the $k_y$ direction. In (b-g), the dispersions (white lines) were extracted from the EDC peak positions (see Supplementary **Figure S2a**), and the blue lines are EDCs obtained at the K point. (h) MDCs at the $E_F$ of the ARPES maps in (b-g).

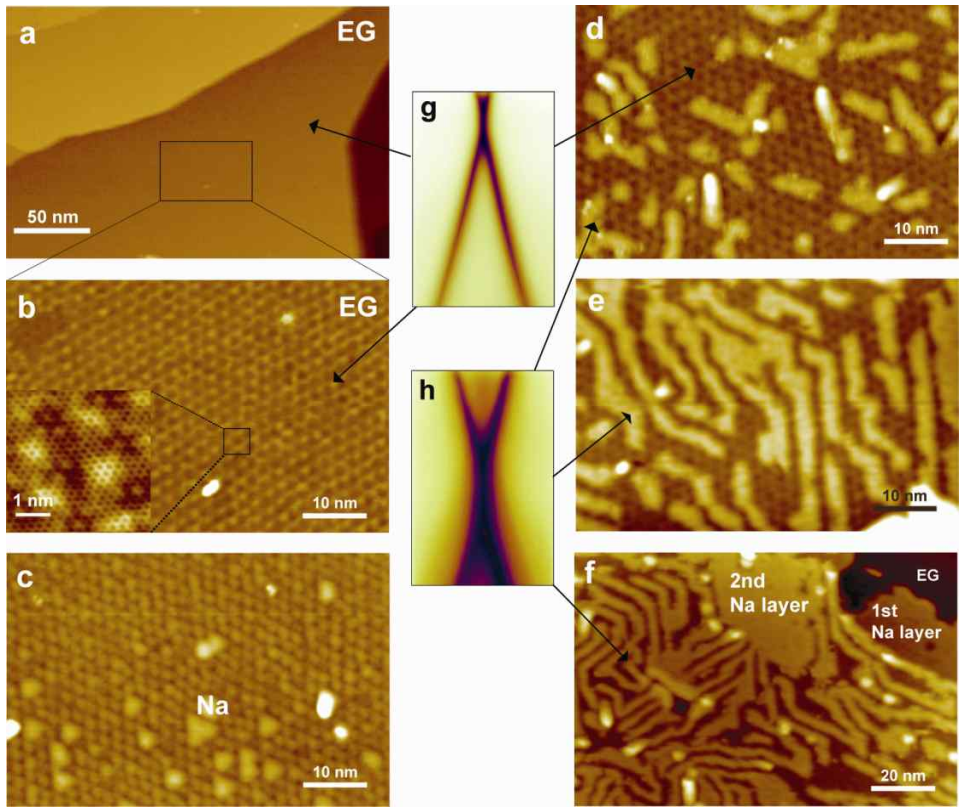

**Figure 2 | Filled-state STM images of monolayer graphene with increasing sodium coverage compared with the ARPES intensity maps**. (a,b) STM image of pristine monolayer graphene, where (b) is the enlarged STM image of the rectangular region in (a). (c-f) STM images of sodium-adsorbed monolayer graphene, where the bright images, which originated from the sodium overlayers, are obviously different from the images of the pristine monolayer graphene. (g,h) ARPES intensity images of pristine (g) and sodium-adsorbed (h) monolayer graphene, where the origins of the STM images of the ARPES intensity maps are indicated by the arrows.

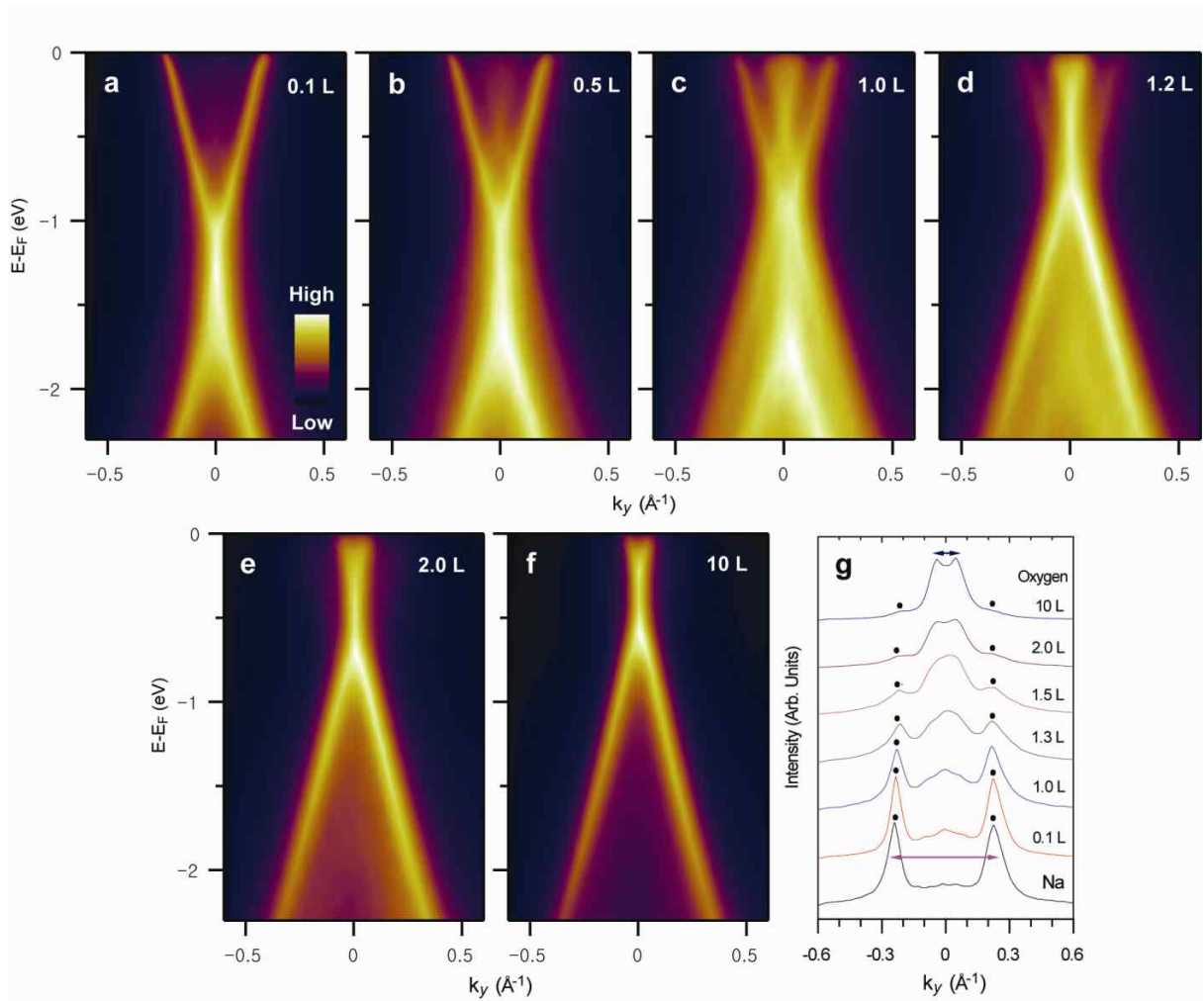

**Figure 3 | Variations in the electronic band structures of the sodium-covered monolayer graphene near the K point with increasing oxygen exposure**. (a) ARPES intensity map of sodium-adsorbed monolayer graphene with saturated sodium coverage. (b-f) ARPES intensity maps measured after the oxidation of the sodium-adsorbed monolayer graphene with saturated sodium coverage. (g) MDCs at $E_F$ of the ARPES maps in (a-f).

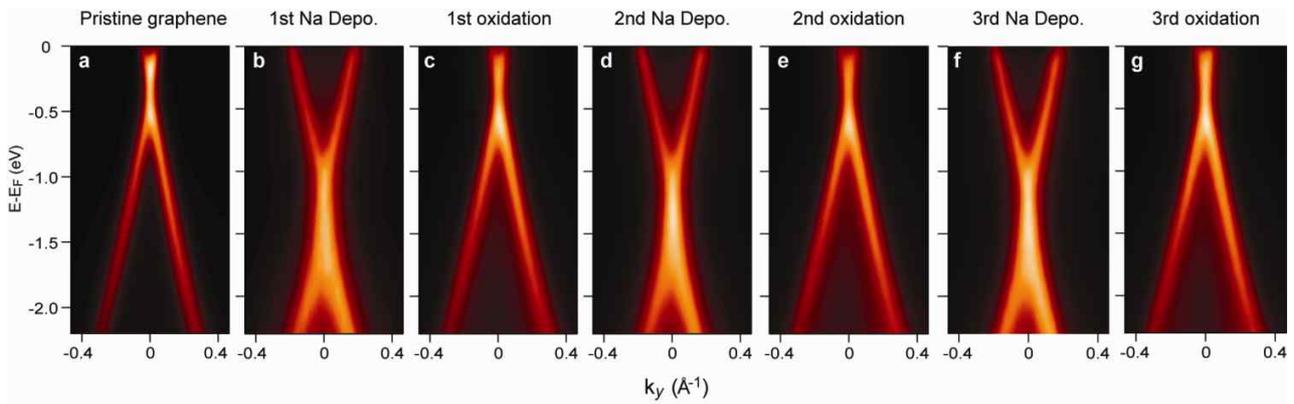

**Figure 4 | Reversible control between the wide-energy-gapped graphene and the pristine graphene by repeated adsorption of sodium and oxygen**. (a) ARPES intensity map of pristine monolayer graphene. ARPES intensity maps measured after the first (b), second (d), and third (f) sodium depositions, where sodium was deposited to saturation coverage. The ARPES intensity maps were measured after the first (c), second (e), and third (g) oxidation, where oxygen was exposed until oxidation was saturated.

# Opening and reversible control of a wide energy gap in uniform monolayer graphene

*Cheolho Jeon[1], Hacheol Shin[1], Inkyung Song[1], Minkook Kim[1], Ji-Hoon Park[1], Jungho Nam[1],*

*Dong-Hwa Oh[1], Sunhee Woo[2], Chan-Cuk Hwang[3], Chong-Yun Park[1] & Joung Real Ahn[1,4]*

[1]Department of Physics, Sungkyunkwan University, Suwon 440-746, Republic of Korea,

[2]College of Pharmacy, Chungnam National University, Daejeon 305-764, Republic of Korea,

[3]Beamline Research Division, PAL, POSTECH, Pohang, 790-784, Republic of Korea,

[4]SAINT and Integrated Nanostructure Physics (CINAP), Institute for Basic Science (IBS),

Sungkyunkwan University, Suwon 440-746, Republic of Korea

Correspondence and request for materials should be addressed to C.Y. Park (cypark@skku.edu) and J.R. Ahn (jrahn@skku.edu).

## S1. Constant-energy maps

**Figure S1** shows the constant-energy maps of the pristine monolayer graphene. The minimum energy gap of monolayer graphene at the Dirac energy is located at the *K* point. Therefore, the determination of the correct position of the *K* point is essential to determine the energy gap at the Dirac energy. The position of the *K* point was determined using the circular constant-energy maps, whose center is the *K* point.

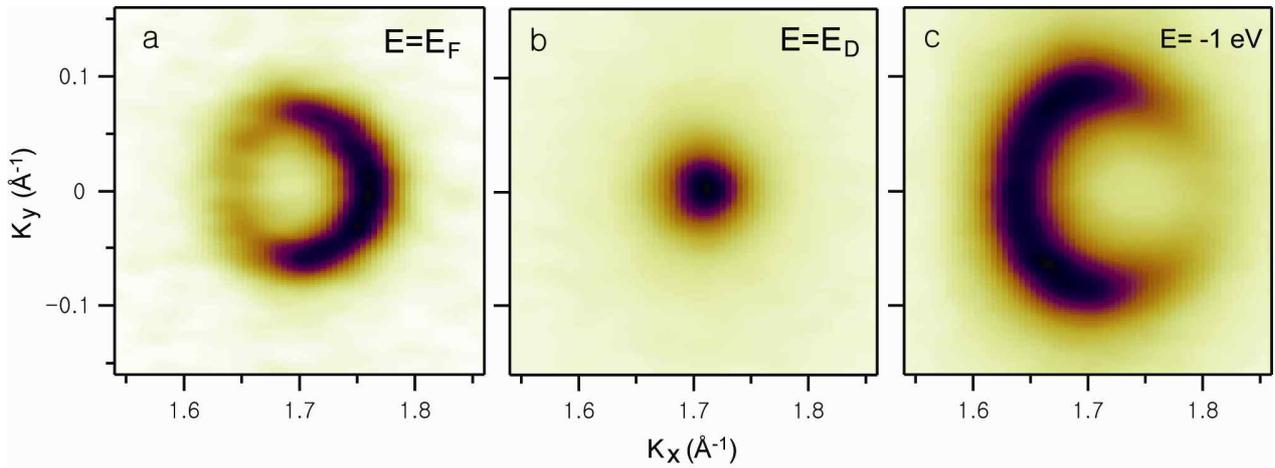

**Figure S1.** Constant-energy maps of the pristine monolayer graphene near the *K* point at (a) the Fermi energy, (b) the Dirac energy, and (c) an energy of -1 eV.

## S2. Energy distribution curves

**Figure S2** shows the energy distribution curves and the ARPES intensity maps of [(a) and (d)] the pristine monolayer graphene, [(b) and (e)] the sodium-covered monolayer graphene with a maximum energy gap of 0.74 eV, and [(c) and (f)] the sodium-covered monolayer graphene with saturation sodium coverage. The minimum energy gap at the Dirac energy could not be determined directly from the ARPES intensity maps. The energy distribution curves were required to determine the exact minimum energy gap, as shown in **Figures S2(a)-(c)**. The red line profiles are the energy distribution curves at the *K* point. The energy gap at the Dirac energy is the energy difference between the two peak positions at each red line profile. The white lines (energy dispersions) in the photoemission intensity maps in **Figures S2(d)-(f)** were extracted from the peak positions of the energy distribution curves.

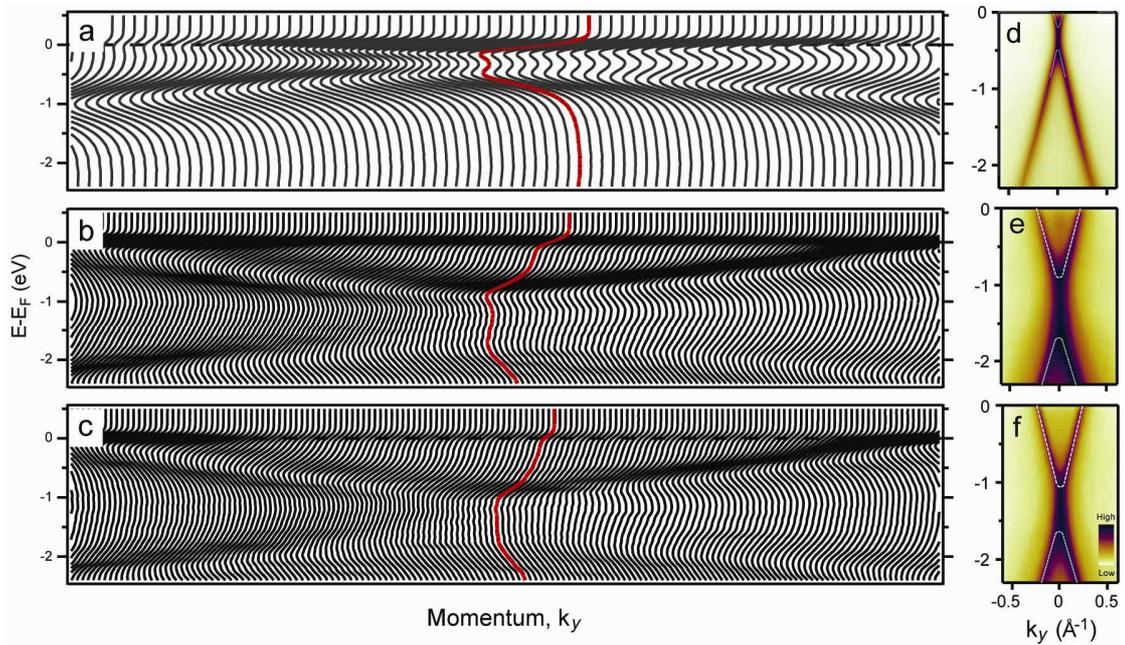

**Figure S2**. (a)-(c) Energy distribution curves and (d)-(f) ARPES intensity maps of the pristine and the sodium-covered monolayer graphene.

**S3. ARPES intensity maps of the sodium-covered monolayer graphene at 60 K**

**Figure S3** shows the ARPES intensity maps of the sodium-covered monolayer graphene along the symmetric direction that crosses the *K* point, where a photon energy of 28 eV was used. After sodium was adsorbed onto the monolayer graphene at room temperature (RT), the sample was cooled to 60 K using liquid helium, and the ARPES intensity maps were measured at 60 K. The ARPES experiment was performed at a low temperature (LT) of 60 K to study the effects of thermal broadening on the ARPES intensity maps measured at RT.

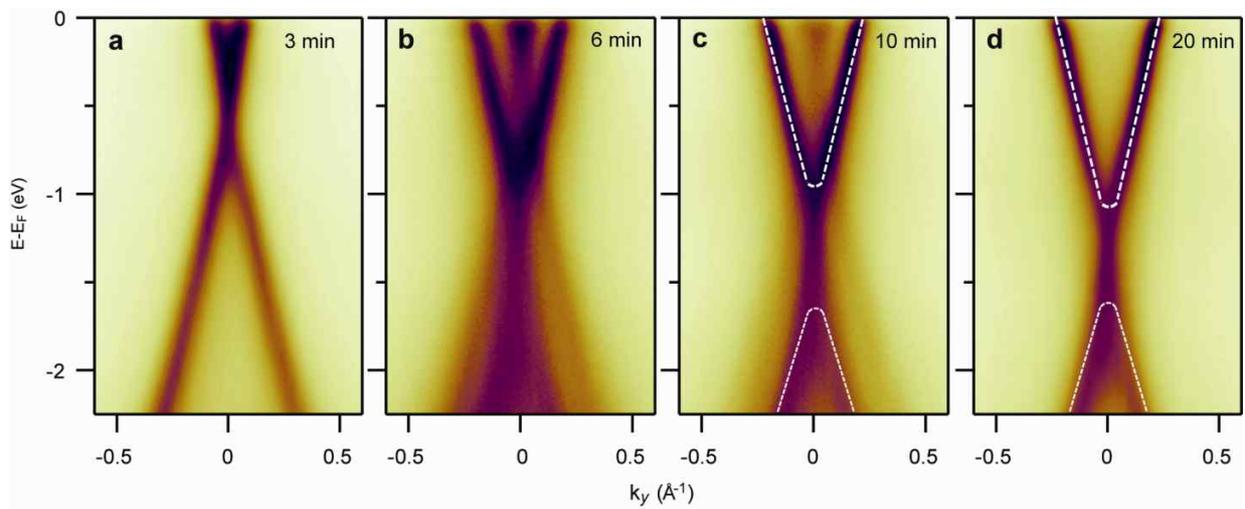

**Figure S3**. (a-d) Variations in the ARPES intensity maps of sodium-covered monolayer graphene near the *K* point were measured at 60 K with increasing sodium coverage, where sodium was adsorbed at RT.

## S4. Adsorption of sodium on monolayer graphene at 60 K

**Figure S4** shows the ARPES intensity maps of sodium-covered monolayer graphene along the symmetric direction that crosses the *K* point, where a photon energy of 28 eV was used. Sodium was adsorbed onto the monolayer graphene at 60 K, and the ARPES intensity maps were measured at the same temperature. The ARPES experiments were performed to compare the changes in the electronic band structure due to sodium adsorption at RT to that at an LT.

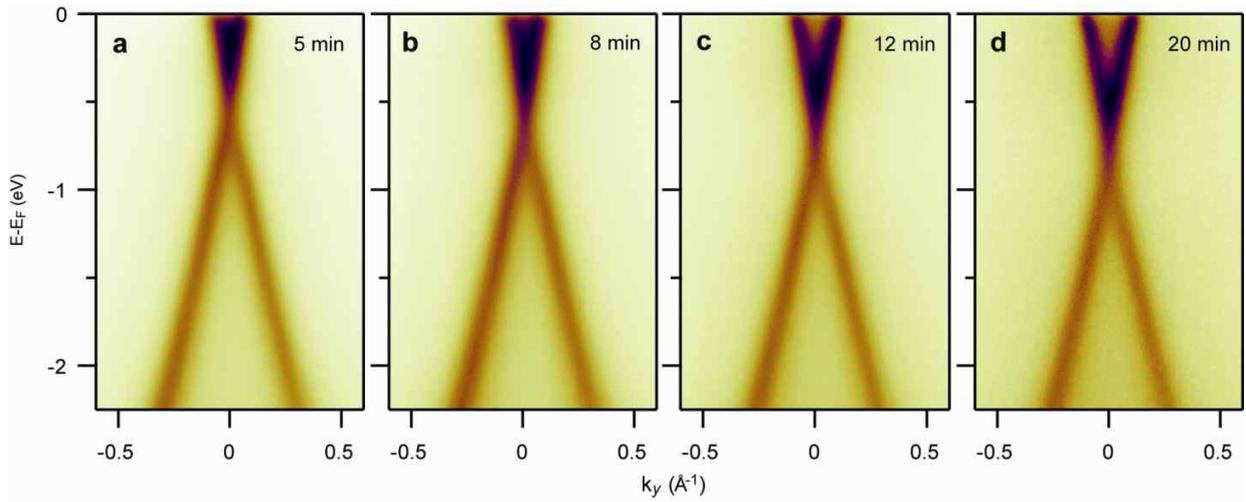

**Figure S4**. (a-d) Variations of the ARPES intensity maps of the sodium-covered monolayer graphene near the *K* point were measured at 60 K with increasing sodium coverage, where sodium was adsorbed at 60 K.

**S5. Adsorption of potassium and cesium on monolayer graphene at RT compared to the adsorption of sodium**

**Figure S5** shows the ARPES intensity maps of (a-d) the sodium-, (e-h) the potassium-, and (i-l) the cesium-covered monolayer graphene along the Γ–K direction, where a photon energy of 40.8 eV was used. The sodium, potassium, and cesium were adsorbed onto the monolayer graphene, and the ARPES intensity maps were at RT. The ARPES experiments were performed to compare the change in an electronic band structure caused by sodium adsorption to that caused by the adsorption of other alkali metals.

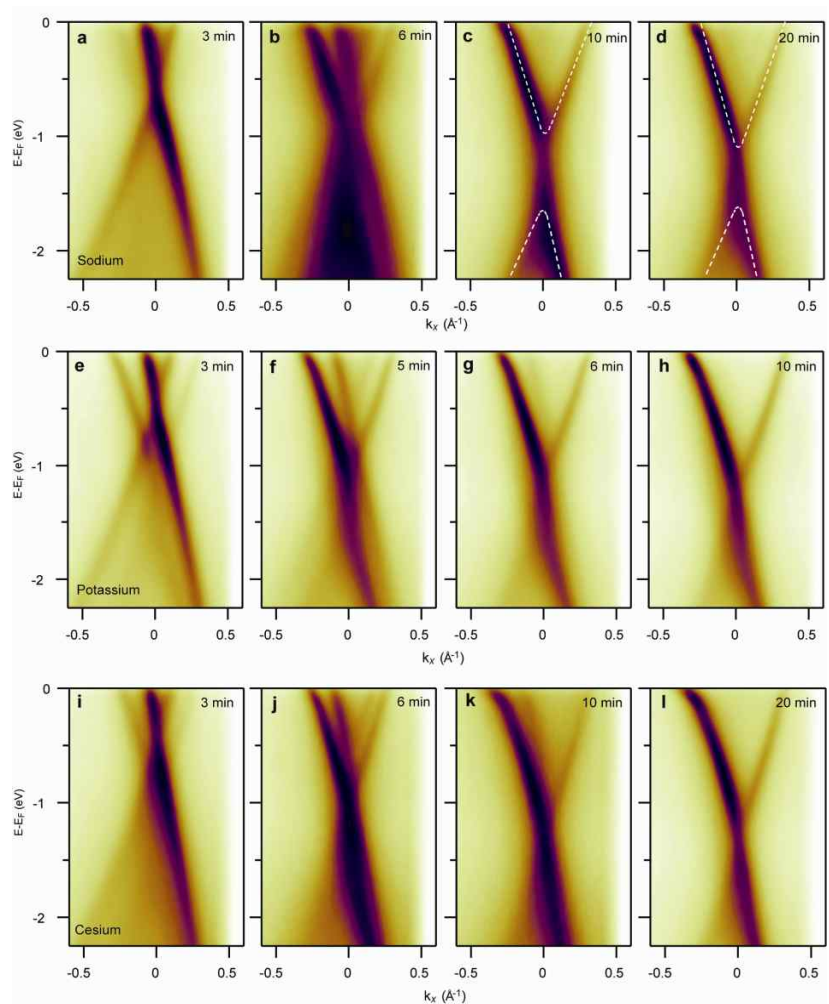

**Figure S5**. Variations in the ARPES intensity maps of (a-d) the sodium-, (e-h) the potassium-, and (i-l) the cesium-covered monolayer graphene along the Γ–K direction were measured at RT with increasing coverage; both the adsorption of alkali metals and the ARPES measurements were performed at RT.

## S6. Change in an LEED pattern with increasing sodium coverage at RT

**Figure S6** shows the LEED patterns of the pristine and the sodium-covered monolayer graphene with increasing sodium coverage at RT. The intensity of the (6√3×6√3)R30° LEED pattern of the pristine monolayer graphene was slightly diminished without the appearance of an additional ordered LEED pattern. This result suggests that, at RT, the Na atoms do not form an ordered structure, such as a (2×2) superstructure.

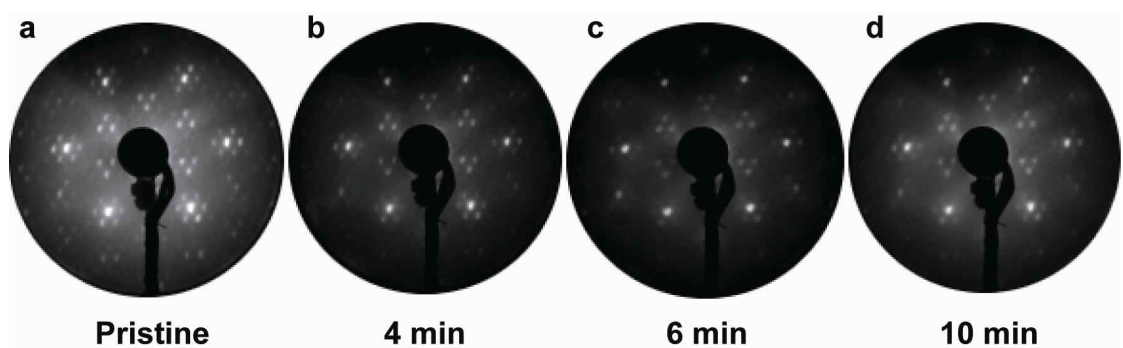

**Figure S6**. LEED patterns of (a) the pristine and (b-d) the sodium-covered monolayer graphene with increasing sodium coverage at RT (beam energy = 100 eV).